\documentclass[12pt]{revtex4}
\usepackage{graphicx}
\usepackage{amsmath}

\begin{document}

\title{Modelling formation and evolution of transverse dune fields}
\author{Jae Hwan Lee$^1$, A. O. Sousa$^1$, E. J. R. Parteli$^1$ and H. J. Herrmann$^{1,2}$}
\affiliation{$^1$Institut f\"ur Computerphysik, Universit\"at Stuttgart, 
	70569 Stuttgart, Germany}
\affiliation{$^2$Departamento de F\'{\i}sica, Universidade Federal do Cear\'a, 
	60455-970, Fortaleza, CE, Brazil}

\date{\today}

\begin{abstract}

We model formation and evolution of transverse dune fields. In the model, only the cross section of the dune is simulated. The only physical variable of relevance is the dune height, from which the dune width and velocity are determined, as well as phenomenological rules for interaction between two dunes of different heights. 
We find that dune fields with no sand on the ground between dunes are unstable, i.e. small dunes leave the higher ones behind. 
We then introduce a saturation length 
to simulate transverse dunes on a sand bed 
and show that this leads to stable dune fields with regular spacing 
and dune heights. 
Finally, 
we show that our model can be used to simulate coastal dune fields if a constant sand influx is considered, where the dune height increases with the distance from the beach, reaching a constant value.

\end{abstract}

\maketitle


\section{introduction}

Sand dunes develop wherever sand is exposed to a medium 
which lifts grains from the ground and entrains them into {\em{saltating}} flow, where grains impacting onto the sand bed eject more and more sand until saturation of the wind is reached leading to a {\em{stationary}} flux. The diverse conditions of wind and sand supply in different regions 
give rise to a large variety of dune shapes.
After the pioneering work by Bagnold \cite{bagnold}, 
dune morphology and dynamics have been investigated many times 
by geologists and geographers
~\cite{finkel_1959,hastenrath_1967,lettau_1969,long,lancaster,tsoar,hesp,wiggs,barbosa,kleinhans}. 
For nearly four decades, 
they have mainly reported field observations about the conditions 
under which the different kinds of dunes appear 
as well as measured their velocity, their shape, the patterns they form, 
the size distribution of sand grains, 
and many other properties. 
However, describing the formation and evolution of a dune field requires 
knowledge of parameters related to the geological history of the area, 
which is in almost all cases not easy to determine. 
Strong variations in sand supply, climate, wind behavior, 
and characteristics of the soil over periods of thousands of years 
are decisive for the actual pattern of the dune fields observed. 

\begin{figure}
\begin{center}
\includegraphics*[width=.55\columnwidth]{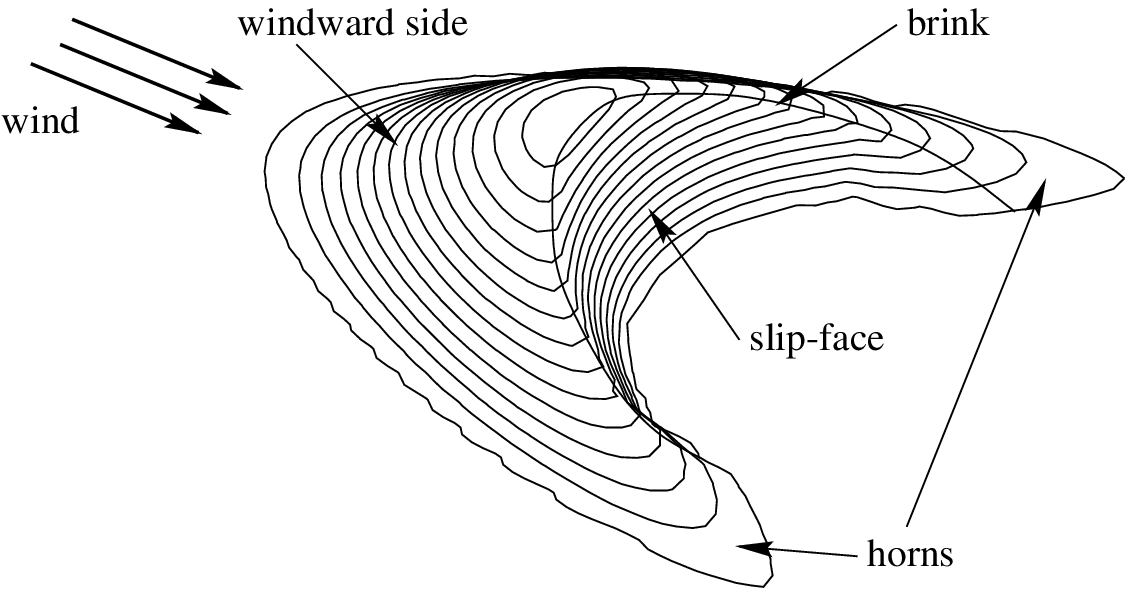}
\includegraphics*[width=0.9\columnwidth]{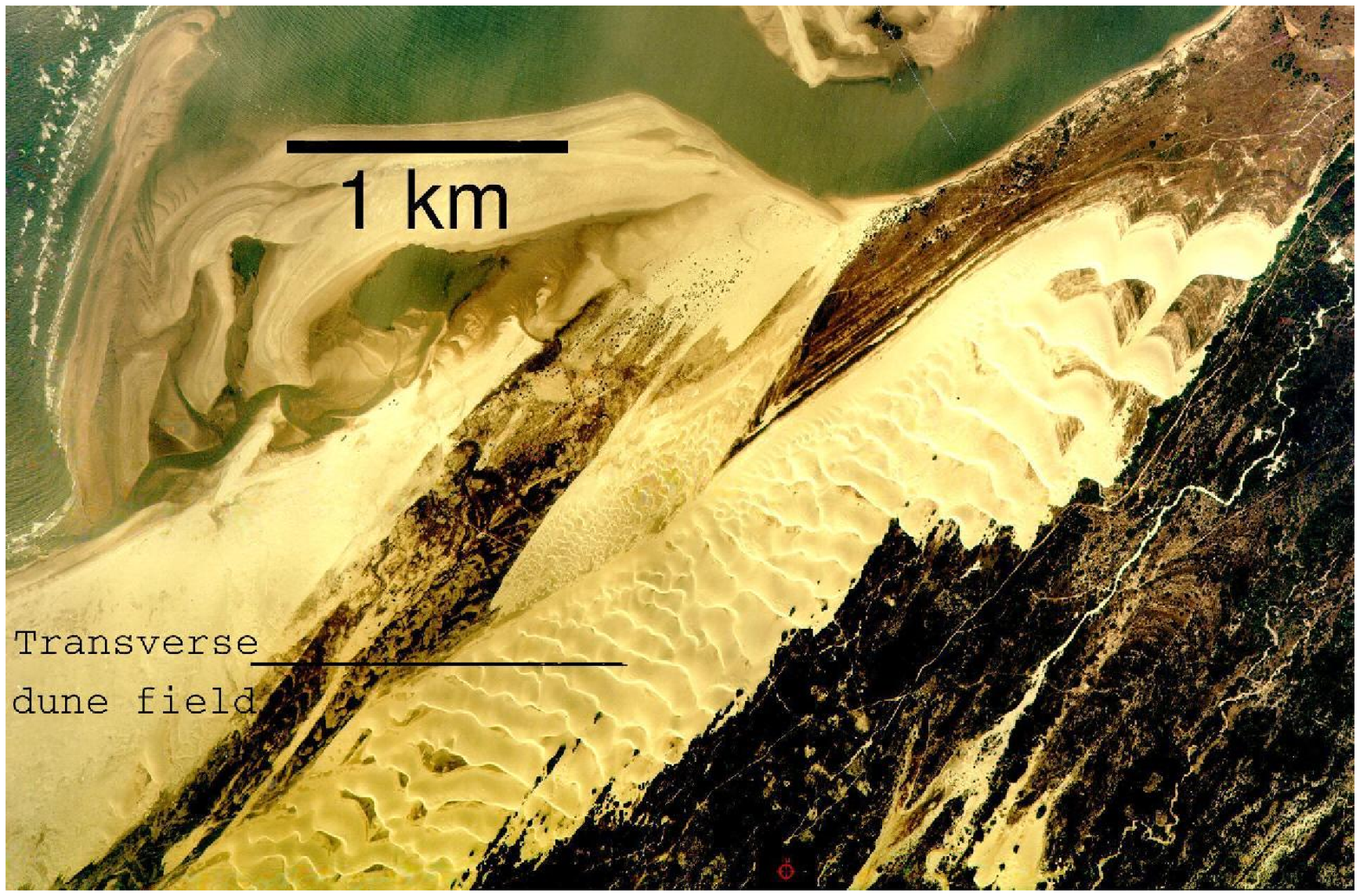}
\caption{Top: Sketch of a barchan dune. Bottom: In the beginning of the dune field known as ``Len\c{c}\'ois Maranhenses'', northeastern Brazil, transverse dunes are formed from the collapse of barchan dunes and extend over more than 20 km downwind. The orientation of the barchanoids indicates the average direction of the wind. Image credit to Embrapa Monitoramento por Sat\'elite, May/1999.}
\label{fig:sketch}
\end{center}
\end{figure}

Almost half of all terrestrial sand seas are covered with transverse dunes \cite{bagnold,lancaster}. 
These dunes appear when unidirectional winds act on areas covered with sand, leading to instabilities of the sand bed \cite{veit_transverse_dunes}, 
or for instance when barchan dunes touch at their horns forming barchanoids \cite{lancaster} and, for increasing amount of sand, giving rise to almost continuous linear chains propagating in the wind direction \cite{parteli_lencois}. In fig.~\ref{fig:sketch} we show the sketch of a barchan dune (top) and an image of the dune field ``Len\c{c}\'ois Maranhenses'' (northeastern Brazil), at its very beginning region, ilustrating a typical situation where an increasing amount of sand yields the transition barchanoids $\rightarrow$ transverse dunes (bottom). 

The mechanisms responsible for transverse dune formation have been extensively investigated by physicists and mathematicians from a theoretical point of view \cite{momiji,andreotti,veit_transverse_dunes}. It has been verified through simulations using a numerical model 
in three dimensions that 
an open system of transverse dunes seems 
to approach lateral translational invariance 
under unidirectional winds \cite{veit_transverse_dunes}, 
allowing modelling transverse dunes in two dimensions. 
In spite of the great theoretical advances, 
many questions concerning the physics of these dunes remain open, 
mainly those related to dune interactions in a field 
and the difficulty in simulating such systems 
$-$ real transverse dune fields may reach lengths of hundreds of kilometers. 

A simple model to study transverse dune fields has been recently introduced~\cite{parteli}. This model is defined in terms of the dune height only, from which sand transport, dune velocity and the field evolution are 
determined through a set of three phenomenological equations. 
The initial conditions consisted of a field of dunes with different heights 
and spacing, 
and the model reproduced the observation 
that in many fields all transverse dunes have similar heights. 
In the present work, we study a different scenario, 
in which the model is used to describe the {\em{formation}} of a dune field. Such scenarios can be found for instance in coastal areas, 
where the sand is transported by the wind from the sea into the continent, 
after being deposited by ocean tides on the beach 
(fig.~\ref{fig:sketch}). 

Furthermore, 
we take into account data of recently performed field measurements 
for the aspect ratio of dunes and for inter-dune distances \cite{parteli_lencois}, which are related to the recirculating flow at the lee side (side of the slip face, where avalanches occur) of the dunes and to the saturation length of the saltating grains \cite{sauermann}. 
Recently, 
Besler~\cite{besler} proposed that dunes may present solitary wave behavior, 
an effect which has been investigated in simulations 
with three-dimensional barchans~\cite{veit_nature,duran}, 
and with transverse dunes in two dimensions~\cite{veit_transverse_dunes}. 
We define dune interaction according to coalescence and solitary wave behavior, where we use relations found for interactions of barchan dunes \cite{duran}, 
since it has been shown that the central transverse profile of symmetric barchans may be a good approximation for the height profile of transverse dunes \cite{sauermann}. 

Our results show that the evolution of dune fields depends strongly on the presence of a sand bed in the field. If there is no sand between dunes, the dune field presents small dunes at its end, which wander with high velocity and increase continuously the size of the field. On the other hand, if the dune field evolves on a sand sheet over which the sand flux is saturated, the smaller dunes do not wander away, and a regular dune spacing is obtained. In this case, we find that the time evolution of the dune spacing in the field agrees well with the results obtained by Werner and Kocurek \cite{kocurek_1999} using a two-dimensional model based on defect dynamics in transverse bedforms. The number of dunes in the field as a function of time is also studied and found to decrease quite regularly in time due to coalescence of dunes. Finally, we use our model to simulate coastal dune fields assuming a constant sand influx downwind.

The model details are presented in the next section. 
In section 3 our results are presented and discussed. Conclusions are made in section 4.


\section{Model description}
\begin{figure}[ht]
\includegraphics*[width=1.0\columnwidth]{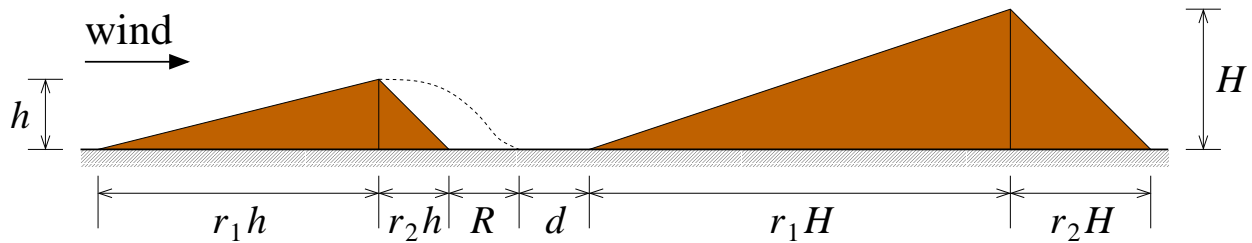}
\caption{Schematic representations of the situation of two transverse dunes.
}
\label{fig:triangles}
\end{figure}
In our model, the variable which determines the field dynamics is the dune height. Transverse dunes move in the $x$ (wind) direction with a velocity $v \equiv dx/dt$ 
inversely proportional to the height $h$ at their crest:
\begin{equation}
v = \frac{a}{h}, 
\label{eq:velocity}
\end{equation}
where $a$ is a phenomenological constant which contains information about wind speed, grain size, etc. \cite{parteli}. A geometrical description of the main elements of our model is shown in fig.~\ref{fig:triangles}, where two dunes of heights $h$ and $H$, $H>h$, have their profiles represented by triangles. The dune width $L$ is related to its height $h$ as $L = (r_1 + r_2)h$, where $r_1 = 10$ and $r_2=1.5$ are constants determined, respectively, by measurements of the height profile at the windward side of transverse dunes in the field \cite{parteli_lencois} and by the fact that the slip face defines an angle of approximately $34^{\circ}$ with the horizontal \cite{bagnold}. The region of recirculating flow at the lee side of the dune, also called ``separation bubble'', has been found to extend downwind from the brink (which in our model coincides with the crest) up to a distance of the order of $2-4$ times the dune height \cite{parteli_lencois}. Here, we use a separation bubble of length $4h$, thus the quantity $R$ in fig.~\ref{fig:triangles} has length $4h - r_2h = 2.5h$ for all dunes. The separation bubble of the dunes has an important implication in interactions of closely spaced dunes, since the net sand flux inside the bubble is zero \cite{sauermann}. The distance measured from the reattachment point of the separation bubble of dune $i$ to the foot of the windward side of the next dune downwind, $i+1$, is called here $d$. 

Dune interaction is defined for the case in which a small dune moving in wind direction behind a larger one eventually reaches, due to its higher velocity, the position of the larger one. In our model, this event is assumed to occur whenever the separation bubble of dune $i$ ``touches'' the windward side of the downwind dune $i+1$, or in other words, when $d<0$ in fig.~\ref{fig:triangles}. It has been found from 3d simulations using a saltation model that barchan dunes may behave as solitary waves \cite{duran}, if their sizes are not too different. The idea is that the interacting dunes exchange sand, the dune downwind decreases in size giving sand to the smaller dune upwind, and they essentially interchange their roles, like if the smaller dune would cross the larger one. However, if the faster dune is significantly smaller than the higher dune downwind, coalescence occurs: The small dune bumps into the larger one and gets ``swallowed up'' \cite{veit_nature,duran}. In our model, dune interaction is defined according to the parameter $\gamma$: If the ratio of the dune volumes ${(V_h/V_H)}_{\mathrm{initial}}={(h/H)}_{\mathrm{initial}}^2$ is smaller than $\gamma$, then the dunes coalesce, and the final dune has height $h_c = {\sqrt{h^2 + H^2}}$ \cite{parteli}; the number of dunes in the field decreases by unity whenever two dunes coalesce. On the other hand, if ${(h/H)}_{{\mathrm{initial}}}^2 > \gamma$, then solitary wave behavior occurs: dunes interchange their roles and the relation between their initial and final heights is as follows \cite{duran}:

\begin{equation}
        \left(\frac{h^2}{H^2}\right)_{\rm final} = C \sqrt{\left(\frac{h^2}{H^2}\right)_{\rm initial} - \gamma},
\label{eq:h2/H2}
\end{equation}
obeying the following mass conservation equation:
\begin{equation}
(h^2 +H^2)_{\rm initial} = (h^2 + H^2)_{\rm final}. \label{eq:mass_conservation}
\end{equation}
Equation (\ref{eq:h2/H2}) is a simplified approximation of the phenomenological observation reported by Duran et. al. \cite{duran} for interactions between barchans, where it is shown that ${(V_h/V_H)}_{\mathrm{final}} \propto \exp{\{-0.22/{[{(V_h/V_H)}_{\mathrm{initial}}-{\mbox{const.}}]}\}}$. Although this has been obtained for barchans, we apply the approach (\ref{eq:h2/H2}) to transverse dunes since careful analysis of 3d simulation results for the shear stress on barchan dunes have shown that the central cut of these dunes may be a good description for the height profile of transverse dunes \cite{sauermann}. While in reference \cite{duran} the value of $\gamma$ that fits the observations for barchans is around $0.15$, in the present work we consider $\gamma$ as a model parameter to study interaction of transverse dunes in the field. The value of the constant $C$ in eq. (\ref{eq:h2/H2}) is determined from a fit to the data by Duran et. al. using the value $\gamma = 0.15$, and is found to be $C \approx 1.3$. We thus identify two limit cases of the parameter $\gamma$: When $\gamma \rightarrow 0$, no coalescence occurs, and the number of dunes in the field remains constant, while for $\gamma \rightarrow 1$ every collision of two dunes results in coalescence and the consequent decrease in the number of dunes in the field. The relations (\ref{eq:h2/H2}) and (\ref{eq:mass_conservation}) are two {\em{independent}} equations that are used to determine the final dune heights after each iteration in which dunes interact like solitary waves.

As mentioned before, the initial condition in all simulations consists of an empty field (i.e. without any dune). Afterwards, transverse dunes with random heights of values uniformly distributed between $h_a$ and $h_b$ are injected into the field at a constant rate $1/{{\Delta}t}$ from the origin $x=0$ downwind, moving with velocity $v(h)$ given by Eq. (\ref{eq:velocity}). The time interval ${\Delta}t$ may vary between 1 and 100 time steps. Typical values of dune heights in the beginning of a dune field are observed to be around $1$m. Thus, to study formation of transverse dune fields, $h_b$ will be set close to $h_a=1$m. However, in cases (i) and (ii) we will use larger values for $h_b$, thus providing a wider spectrum of dune heights in order to study the effect of coalescence and solitary wave behavior in the evolution of a dune field. In our simulations, each time step is defined as $0.01$ year. Thus, after each iteration dunes move a distance downwind given by ${(0.01 \times a)}/h$ m, and may interact with each other according to the rules mentioned before. The typical simulation time for 10000 iterations is a few tens of seconds running on a Pentium IV.

Following the dynamics described above, we study three cases: (i) Dunes are injected into a field with no sand on the ground; (ii) Dune fields are studied when the ground is covered with a sand bed, where a phenomenological parameter, the saturation length of the saltation sheet, is introduced. (iii) Finally, we add to case (ii) a differential equation to simulate a sand influx, which changes the heights of the dunes over the field evolution. 


\section{Results and Discussion}

In cases (i) and (ii) we investigate the evolution of the field due only to interactions of dunes through coalescence and solitary wave behavior, according to relations ({\ref{eq:h2/H2}}) and (\ref{eq:mass_conservation}), which are the mechanisms leading to changes in the heights of the dunes. In this case, the mean dune height in the field doesn't go beyond the maximum height $h_b$ of the injected dunes. 

\begin{figure}[!ht]
\begin{center}
 \includegraphics*[width=.85\columnwidth]{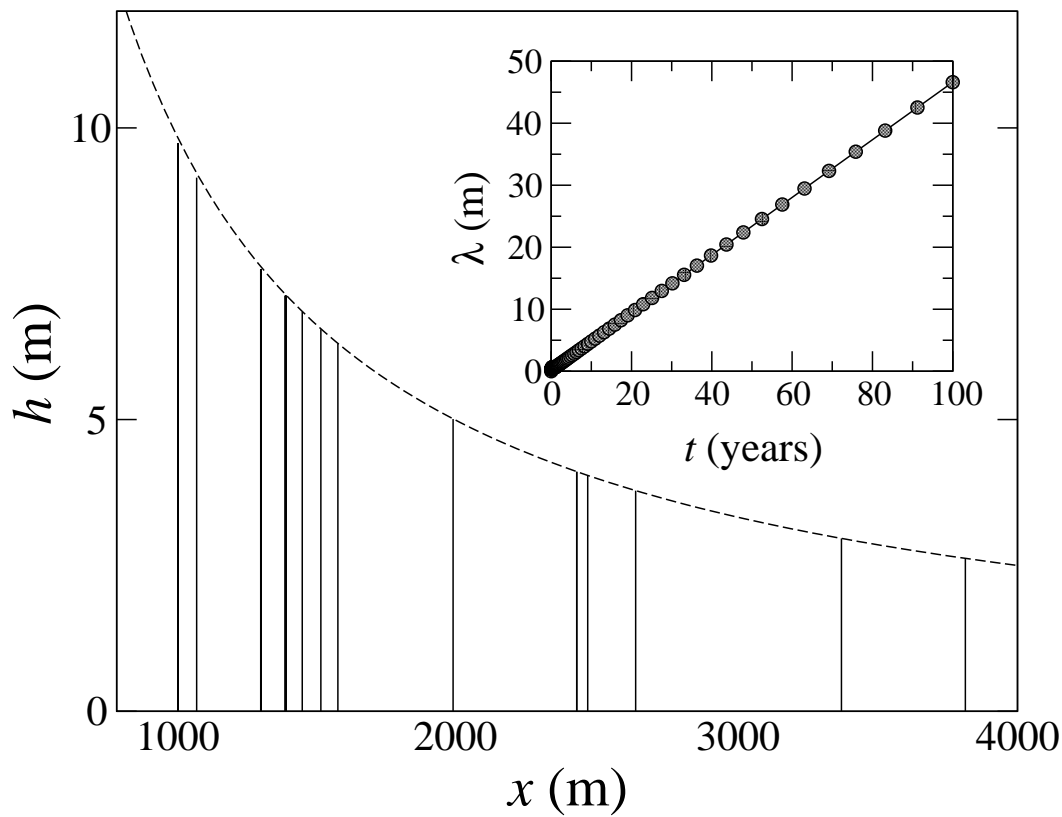}
\caption{The main plot shows the profile of a simulated dune field after $t=100$ years, obtained after input of 100 dunes from wind direction. Dune crests are represented by the vertical line segments. The dashed line represents the curve $h=10^4/x$. Model parameters are $a=100$ m$^2/$year, ${\Delta}t= 0.01$ year and $\gamma = 0.234$. The inset shows the average dune spacing, representing ${\lambda} \simeq 0.46\,t$.} 
\label{fig:field_without_ls}
\end{center}
\end{figure}

Figure \ref{fig:field_without_ls} shows the snapshot of a transverse dune field at $t=10^4$ time steps ($100$ years) for case (i) when 100 dunes are injected into the field at regular time intervals ${\Delta}t= 0.01$ year, using $a=100$ m$^2/$year, $h_b=10$m, and $\gamma = 0.234$. Each vertical line segment in this plot corresponds to the height at the crest of a single dune.
Notice that $h_b=10$m is a value quite unrealistic to study the origin of a dune field where dunes are injected from the beach. However, as mentioned above, we are in this first part of our work only interested in studying the effect of interactions between dunes during the field evolution. Figure \ref{fig:field_without_ls} shows that the initial number of dunes has been reduced by coalescence. Furthermore, we can see that the height of the dunes in the field decreases as the inverse of the dune position downwind: The smaller dunes are found at the end of the field. The dashed line in the main plot represents the curve $h = 10^4/x$. The prefactor $A$ of the curve $h(x) = A/x$ is time dependent, and may be determined from Eq. (\ref{eq:velocity}), which leads to $h(x) = at/x$. The reason for this behavior is that small dunes bump into larger ones giving rise to the large height profile at the beginning of the field, while dunes interacting like solitary waves contribute to the emergence of the small dune heights observed at larger values of $x$, and the consequent temporal increase of the field length, i.e. the distance between the dunes at both extremities of the field. We have also observed that the curve $h(x)=A/x$ is independent of the value of $\gamma$. The inset of fig.~\ref{fig:field_without_ls} shows that the mean spacing $\lambda$ between the dunes increases indefinitely with time, since the small dunes at the end of the field, due to their higher velocities, get more and more apart from the larger ones. Notice that the average dune spacing $\lambda$ for one particular realization of transverse dune field may be written as a linear function of the time $t$ with coefficient in terms of the heights of the first and last dunes, $h_1$ and $h_N$, respectively, as well as of the total number of dunes in the field, $N$:
\begin{equation}
\lambda = \frac{a}{N-1} \left({{\frac{1}{h_N}} - \frac{1}{h_1}}\right)t \label{eq:spacing}.
\end{equation}
For the dune field shown in fig.~\ref{fig:field_without_ls}, we have $N=61$, $h_1=2.6$m and $h_2=9.7$m, then we find that the coefficient of Eq. (\ref{eq:spacing}) is approximately $0.46$, which agrees with the slope of the curve shown in the inset of this figure. 

In summary, our findings for case (i) show transverse dune fields with decreasing dune heights with distance and an increasing dune spacing with time, since the smaller faster dunes wander at the end of the field. However, field measurements \cite{Lancaster_1983} have shown that dune spacing {\em{increases}} with the dune height, as opposed to the situation found in fig.~\ref{fig:field_without_ls}, where spacing between smaller dunes is mostly larger than for higher dunes. Furthermore, one relevant aspect to be noticed in the dune field of fig.~\ref{fig:field_without_ls} is the quite irregular dune spacing, where ``clusters'' of dunes are even observed, which are not found in real transverse dune fields.


An important difference between transverse dune fields observed on terrestrial sand seas and the dune fields simulated so far using our model lies in the availability of sand on the ground. In dune fields where transverse dunes appear on a sand sheet over which the sand flux between the dunes is saturated, regular spacing and similar dune heights are observed~\cite{parteli_lencois}. This happens because the availability of sand yields a saturated flux between dunes. Saturation is reached within a time interval of a few seconds, and is associated with the saturation length $l_s$, which is found to decrease with the strength of the wind \cite{sauermann}. In a transverse dune field developed on a sand bed, this length may be measured from the reattachment point of the separation bubble of the upwind dune (see fig.~\ref{fig:triangles}), where the wind strength increases from zero (value of the wind strength in the separation bubble) and sand transport initiates \cite{sauermann}. When the flux becomes saturated, i.e. after a distance $d=l_s$ from the reattachment point, sand is deposited at the foot of the windward side where the wind strength decreases, and the surface at the beginning of the downwind dune is not eroded. It follows that no dune can move faster than its upwind neighbor. 

To simulate such dune fields, we introduce in our model the phenomenological constant $l_s$ by fixing for the quantity $d$ in fig.~\ref{fig:triangles} a maximum value $d_{\mathrm{max}}=l_s$. We set $l_s=2$ m. As a consequence, the {\em{maximum}} crest-to-crest distance in the field, $D_{\mathrm{cc}}$, is calculated as:
\begin{equation}
D_{\mathrm{cc}} = r_2h + R + l_s + r_1H, 
\label{eq:Dcc}
\end{equation}
according to the definitions in fig.~\ref{fig:triangles}. 
It is important to notice that the maximum spacing $D_{\mathrm{cc}}$ in the field, as predicted by Eq. (\ref{eq:Dcc}), is not a constant, but does depend on the dune size, and is larger in dune fields of higher dunes. 

\begin{figure}
\begin{center}
\includegraphics*[width=.9\columnwidth]{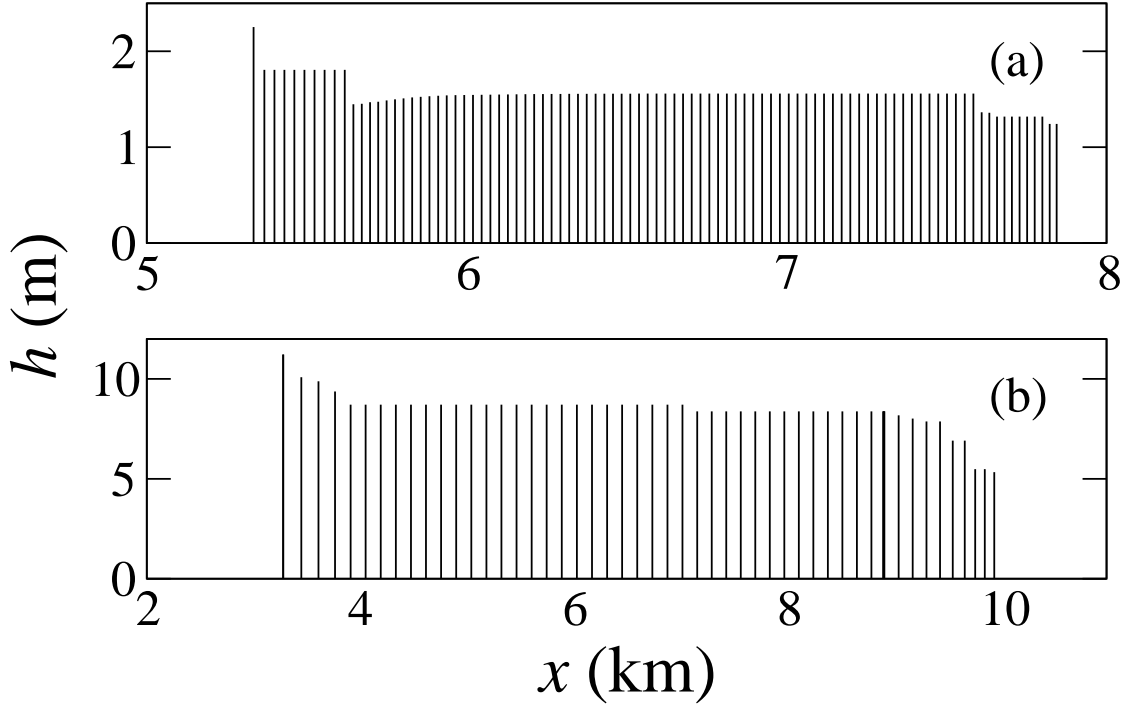}
\caption{Transverse dune field obtained when the saturation length $l_s$ of the saltation sheet is considered. 
Model parameters are ${\Delta}t = 0.1$ year, $a=150$ m$^2/$year, $\gamma = 0.30$, 
and (a) $t=90$ years (b) $t=500$ years. 
The heights of the 100 injected dunes at $t=0$ are random numbers 
between 1 and (a) 2m, (b) 10 m. 
We see that the two fields present a quite regular dune spacing.}
\label{fig:field_with_ls}
\end{center}
\end{figure}

Figure \ref{fig:field_with_ls} shows two simulated dune fields when the saturation length $l_s$ is taken into account (case (ii)). The profile shown in fig.~\ref{fig:field_with_ls}(a) corresponds to $t=90$ years, where dunes of initial heights between $h_a=1$m and $h_b=2$m have been injected into the field (now a ground covered with sand) using ${\Delta}t = 0.1$ year, $a=150$ m$^2/$year and $\gamma = 0.3$. As we can see in this figure, all transverse dunes reach approximately the same height, around $1.6$m, and present a quite regular spacing. Figure~\ref{fig:field_with_ls}(b) shows another dune field obtained at $t=500$ years with $h_a=1$m and $h_b=10$m, ${\Delta}t = 0.1$ year, $a=150$ m$^2/$year and $\gamma = 0.3$, where we find dunes regularly spaced and of heights around 8m. As we can see, the value of the final dune height depends on the initial values of $h_a$ and $h_b$. Moreover, comparison of fig.~\ref{fig:field_with_ls} with fig.~\ref{fig:field_without_ls} shows that the introduction of the saturation length leads to much more stable transverse dune fields, since the smaller dunes wandering fast at the end of the field are now not found anymore. 

\begin{figure}
\begin{center}
  \includegraphics*[width=.7\columnwidth]{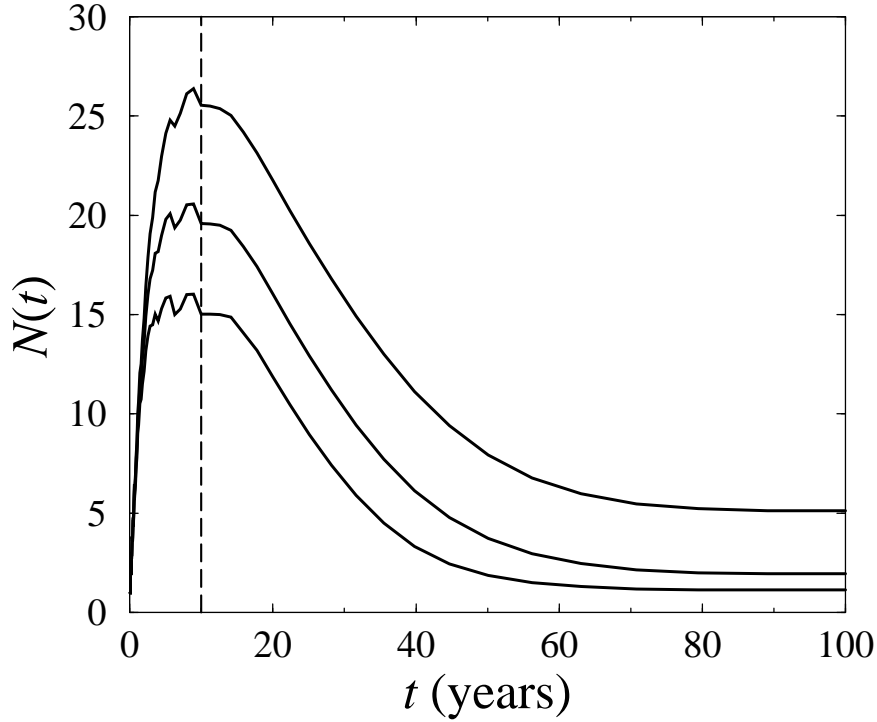}
\caption{Number of dunes in the field, $N(t)$, 
as a function of time $t$ for $\gamma = 0.34$, $0.35$ and 
$0.36$ (from top to bottom) averaged over 1000 realizations. The dashed line 
indicates the time when input of the 100 injected dunes with random heights between 1 and 2 meters finishes. ${\Delta}t=0.1$ year and $a=300$ m$^2/$year.}
\label{fig:N(t)}
\end{center}
\end{figure}

\begin{figure}
\begin{center}
\includegraphics*[width=.7\columnwidth]{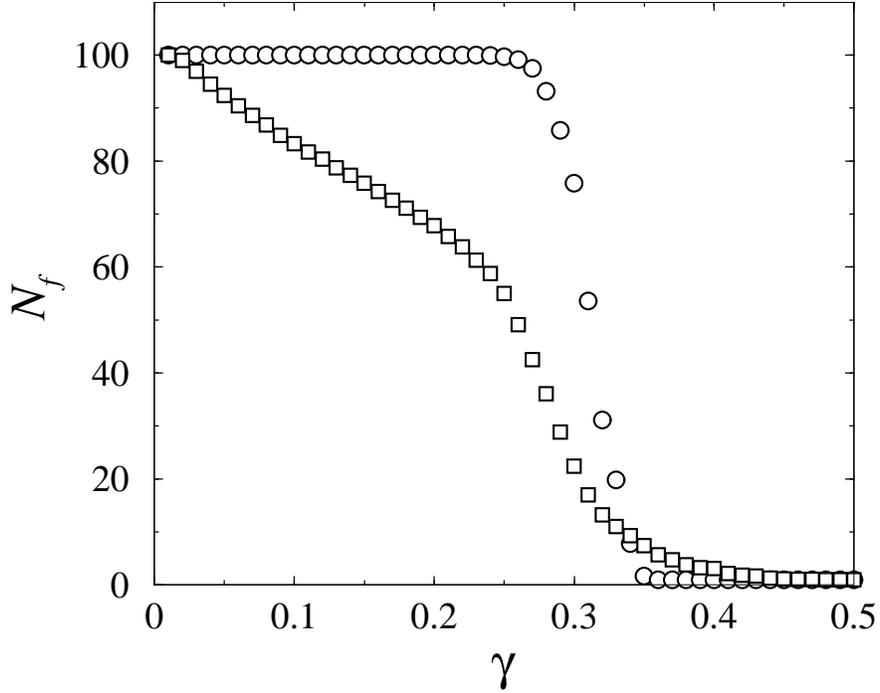}
\caption{Final number of dunes $N_f$ as a function of $\gamma$, averaged over 1000 realizations. The heights of the 100 injected dunes are random numbers between 1 m and $h_b$, 
where $h_b=2$ m (circles) and $10$ m (squares). 
The number of dunes decreases in time due to coalescence between dunes. ${\Delta}t=0.1$ year and $a=300$ m$^2/$year.}  
\label{fig:N_f}
\end{center}
\end{figure}

The time evolution of the number of dunes in the field is shown in fig.~\ref{fig:N(t)} for three different values of the parameter $\gamma$, using again ${\Delta}t=0.1$ year and $a=300$ m$^2/$year. As we can see in this figure, after injection of the 100 dunes into the field (at $t=10$ years as indicated by the dashed line), the number of dunes decreases quite regularly in time, as also found by Schw\"ammle et. al. \cite{veit_transverse_dunes} from a two-dimensional model of transverse dune fields. Furthermore, after a certain time, when dunes reach the same height, the number of dunes gets constant, $N_f$. In fig.~\ref{fig:N_f}, we present the final number of dunes $N_f$ as a function of $\gamma$, averaged over 1000 realizations, when $h_b = 2$m (circles) and $h_b=10$m (squares), using $h_a=1$m, ${\Delta}t=0.1$ year and $a=300$m$^2/$year. As we can see in this figure, for small values of $\gamma$ ($\gamma \rightarrow 0$), the number of dunes remains at the initial value $N=100$, since coalescence occurs less frequently. In particular, this effect is more visible when $h_b=2$ m (circles), because in this case dunes have more similar initial heights (between 1m and 2m), and coalescence between dunes occurs even more rarely. As the value of $\gamma$ increases, $N_f$ decreases because dunes coalesce more often in the field; for large values of $\gamma$, as discussed before, $N_f$ is equal to 1. One aspect of particular interest in fig.~\ref{fig:N_f} is that $N_f$ decreases quite suddenly with $\gamma$ if the initial heights of the dunes are very similar (curve represented by circles). In particular, we have found that as $h_b \rightarrow h_a=1$m, $N_f$ decreases rapidly to unity at $\gamma \approx 0.30$. On the other side, for a wider range of heights of the injected dunes ($h_b \gg h_a$), $N_f$ decreases more smoothly with $\gamma$, since in this case even small values of $\gamma$ are large enough for coalescence to occur.
%
%

\begin{figure}
\begin{center}
\includegraphics*[width=.75\columnwidth]{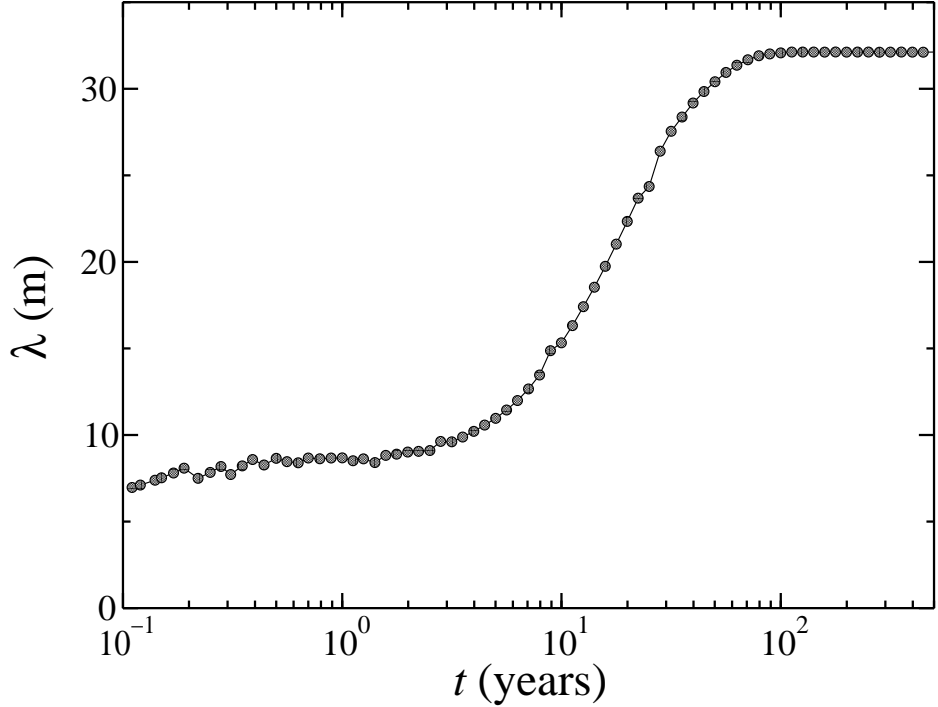}
\caption{Mean dune spacing $\lambda$ as a function of time, averaged over 1000 realizations of dune fields using $\gamma = 0.3$, $a=100$ m$^2/$year and ${\Delta}t=0.1$ year, after injection of the 300 dunes, which have initial random heights between 1 m and $h_b=2$ m. Computational time is 300 minutes on a Pentium IV computer to obtain this figure.}
\label{fig:lambda}
\end{center}
\end{figure}

Figure \ref{fig:lambda} shows the average dune spacing $\lambda$ as a function of time using ${\Delta}t=0.1$ year, $a=100$ m$^2/$year, $\gamma = 0.3$, and considering the saturation length $l_s=2$m.  
As we can see from comparison of this figure with the inset of fig.~\ref{fig:field_without_ls}, the inter-dune distances evolve in time in a very different way as compared to the case where the saturation length $l_s$ was not taken into account. Figure \ref{fig:lambda} shows that the dune spacing now increases until it reaches a {\em{saturation value}}, which corresponds to the equilibrium state of the field, where dunes have similar heights as shown in fig.~\ref{fig:field_with_ls}. 

A very similar result was found recently by Werner and Kocurek \cite{kocurek_1999} using a transverse dune model in two dimensions based on defect dynamics. They found that $\lambda$ increases logarithmically in an intermediate time interval until getting constant, as observed in fig.~\ref{fig:lambda}. Their model consists of a set of differential equations to describe the dynamics of the endlines of two-dimensional transverse dunes (length and width). In their model, which was used to predict ages of real dune fields \cite{kocurek_1999}, interaction between dunes led to a decrease in the number of defects in the field. We would like to remark that, although fig.~\ref{fig:lambda} suggests that ${\lambda}(t)$ increases approximately with ${\log}(t)$ for $8 \leq t \leq 40$, we could not find more than one decade of this possible logarithmic behavior of dune spacing in time.

It is important to notice that the model defined in case (ii) in the absence of any sand influx reproduced formation of dune fields of regular spacing, just by injection of dunes downwind onto an empty ground, which interact by coalescence or solitary wave behavior according to relations (\ref{eq:h2/H2}) and (\ref{eq:mass_conservation}). This situation has to be differentiated from the case studied in ref. \cite{parteli}, where a constant sand influx is considered, and a field of transverse dunes of random initial heights evolves in time until a constant final height is reached, which is proportional to the sand influx. There, a regular spacing was only found in the case where the initial dune spacing and heights were considered to be also regular. Of course, a serious limitation of our model so far is that the final height of the dunes is limited according to their initial values. 

\begin{figure}[ht]
\begin{center}
\includegraphics*[width=0.95\columnwidth]{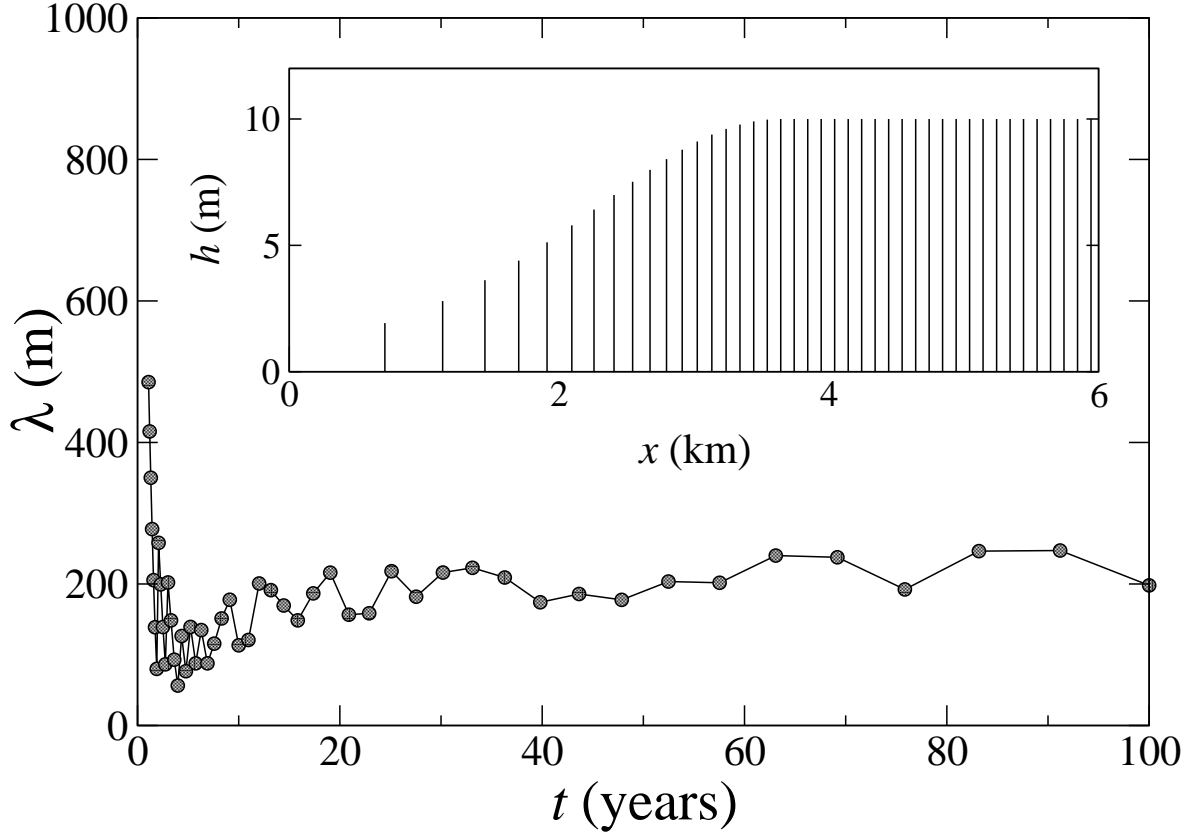}
\caption{The inset shows a simulated dune field after $t=100$ years, 
using the constant influx according to Eq. (\ref{eq:dh/dt}). 
Model parameters are $a=1000$ m$^2/$year, $b=10^3$ year$^{-1}$, $c=0.01$, ${\phi}_0 = 1.0 \times 10^4$ m$/$year, $\gamma=0.3$ and ${\Delta}{t}=1.0$ year. 
The main plot shows the dune spacing $\lambda$ as a function of time averaged over 1000 realizations.} 
\label{fig:field_with_ls_with_influx}
\end{center}
\end{figure}

We now define the following equation for the dune height:
\begin{equation}
\frac{dh_i}{dt} = c \times {\sqrt{|{\phi}_0-{\phi_{i}}|}},
\label{eq:dh/dt}
\end{equation}
where ${\phi}_0$ is the constant influx, $\phi_i=bh_i$ is the sand flux at the position of dune $i$ \cite{parteli}, and $b$ and $c$ are phenomenological constants which depend on the conditions of the field as wind velocity, and sand supply. It is important to notice that Eq. (\ref{eq:dh/dt}) does not include sand transport between dunes, as opposed to the case studied in ref. \cite{parteli}. We follow the same rules for dune interaction as mentioned before and consider transverse dunes evolving on a sand sheet, where the saturation length $l_s$ is taken into account as in the previous case. In the inset of fig.~\ref{fig:field_with_ls_with_influx}, we show a typical transverse dune field obtained after $t=100$ years using Eq. (\ref{eq:dh/dt}) and injecting dunes with heights $h_a=h_b=1$m at rate $1/{{\Delta}t}$, where ${\Delta}t=1.0$ year. The parameters used are $a=1000$ m$^2/$year, $b=10^3$ year$^{-1}$, $c=0.01$, ${\phi}_0 = 1.0 \times 10^4$ m$/$year, and $\gamma=0.3$. As we can see, in the beginning of the field, dune heights {\em{increase}} with the position downwind. Furthermore, some kilometers later, dunes reach regular spacing and similar heights $h_0 = {\phi}_0 /b$ \cite{parteli}. The scenario presented in fig.~\ref{fig:field_with_ls_with_influx} corresponds to an {\em{evolving}} field where dunes are continuously injected; it is no equilibrium state. Such a picture can be found for instance in many parts of the coastal dune field of the Len\c{c}\'ois Maranhenses. Therefore, our model can reproduce dune fields observed on the coast, where new dunes are continuously appearing in the beginning of the field, at the beach. 

\begin{figure}
\begin{center}
\includegraphics*[width=.7\columnwidth]{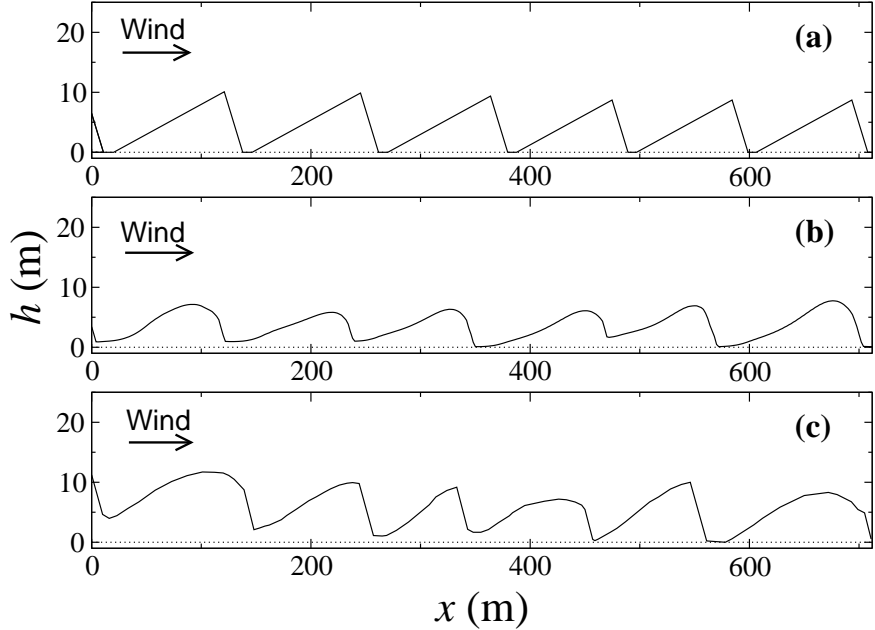}
\caption{In (a) we show a typical dune field obtained with our model where the saturation length $l_s$ is considered, with $a=300$ m$^2/$year, ${\Delta}t = 0.1$ year and $\gamma=0.3$, $N=100$ injected dunes and $t=1000$ years. (b) and (c) are figures of reference \cite{parteli_lencois}; they are respectively a simulation of transverse dunes in two-dimensions and the measured profile of a transverse dune field in the Len\c{c}\'ois Maranhenses (fig.~\ref{fig:sketch}).}
\label{fig:comparison}
\end{center}
\end{figure}
The main plot of fig.~\ref{fig:field_with_ls_with_influx} shows the dune spacing as function of time averaged over 1000 realizations. It can be seen that, after a certain number of time steps, the dune spacing reaches a constant value corresponding to the inter-dune distances for which dunes present the equilibrium height of $10$ m. In fig.~\ref{fig:comparison}, we compare the dune field obtained from our simulations (fig.~\ref{fig:comparison}(a)) with the profile of the measured dunes in the Len\c{c}\'ois Maranhenses (fig.~\ref{fig:comparison}(c)) and simulation results from a two-dimensional model of closely spaced transverse dunes (fig.~\ref{fig:comparison}(b)) \cite{parteli_lencois}. The height profile of the dunes shown in (a) follows the relations defined in fig.~\ref{fig:triangles}. 
 
In summary, based on our results, we have shown that the sand flux relation (\ref{eq:dh/dt}) is of fundamental importance to reproduce realistic values of dune heights observed in real transverse dune fields from realistic values of initial dune heights. Without Eq. (\ref{eq:dh/dt}), dunes with small initial heights in the beginning of the field cannot grow to the heights observed in real dune fields, as can be seen in figs. \ref{fig:field_with_ls}(a) and \ref{fig:field_with_ls}(b) where the final dune height is always around $h_b$. Another aspect to be mentioned is the calculation of the parameter $\gamma$ for transverse dunes. Due to the great simplicity of our model, no conclusion could be made about a single threshold value $\gamma = {\gamma}_{\mathrm{transverse}}$ which would describe interactions of transverse dunes, as done recently for barchans in 3 dimensions, where ${\gamma}_{\mathrm{barchan}} \approx 0.15$ \cite{duran}. To this point, simulations of transverse dunes in two dimensions would be required. It would be also interesting to check if the results found in figs. \ref{fig:N_f}, \ref{fig:lambda} and \ref{fig:field_with_ls_with_influx} may be found from other numerical models of transverse dune fields in two and/or three dimensions. 


\newpage

\section{Conclusions}

We presented a simple model to study formation of transverse dune fields. The model was used to simulate the evolution of dune fields where neighboring dunes may coalesce or exhibit solitary wave behavior according to their relative volumes. We introduced a phenomenological parameter, $\gamma$, which was compared to the ratio $h^2/H^2$ of the volumes of the interacting dunes of heights $h$ and $H$, $h<H$, where coalescence and solitary wave behavior were defined respectively for $h^2/H^2 < \gamma$ and $h^2/H^2 > \gamma$. We found that this simple rule led to a strong dependence of the field evolution on $\gamma$. We have shown that dune fields formed by input of dunes onto an empty field with no sand on the ground present decreasing dune heights and increasing spacing with the distance downwind, and the length of the field increases indefinitely with time. On the other hand, we found that the dune height and spacing reach an equilibrium value if the saturation length $l_s$ of the saltation sheet is considered, where we simulated transverse dune fields evolving on a sand bed. However, the final dune heights were limited due to the absence of a sand influx. By introducing a sand influx, we could simulate coastal dune fields with crescent dune heights with distance. We have shown that, in spite of the simplicity of the model, our results agree with predictions of simulations from more complex dune models in two dimensions and reproduce well real transverse dune fields. 

\acknowledgments

We acknowledge O. Dur\'an for very important suggestions and discussions, and V. Schatz for a critical reading of this manuscript. This work was supported in part by the Max-Planck Prize awarded to H. J. Herrmann (2002). E. J. R. Parteli acknowledges support from CAPES - Bras\'{\i}lia/Brazil.


\newpage

\end{document}